\documentclass[]{spieman}

\usepackage[utf8]{inputenc}
\usepackage{amsmath,amsfonts,amssymb}
\usepackage{graphicx} 
\usepackage{grffile} 
\usepackage{hyperref} 
\usepackage[symbol]{footmisc}
\usepackage{rotating}

\title {The WIYN One Degree Imager in 2018: An Extended 30-Detector Focal Plane}

\author[a,b]{Daniel R. Harbeck} \author[c]{Mike Lesser} \author[b]{Wilson Liu}
\author[d]{Bob Stupak} \author[d]{Ron George} \author[d]{Ron Harris}
\author[d]{Gary Poczulp} \author[d]{Jayadev Rajagopal} \author[e]{Ralf Kotulla}
\author[c]{David Ouellete} \author[b, e]{Eric J. Hooper} \author[e]{Michael Smith}
\author[e]{Dustin Mason} \author[f]{Peter Onaka} \author[f]{Greg Chin}
\author[d]{Emily Hunting} \author[d]{Robert Christensen}

\affil[a]{Las Cumbres Observatory, Goleta, CA (USA)} \affil[b]{WIYN Observatory,
Tucson, AZ (USA)} \affil[c]{The University of Arizona, Tucson, AZ (USA)}
\affil[d]{NOAO, Tucson, AZ (USA)} \affil[e]{University of Wisconsin, Madison, WI
(USA)} \affil[f]{University of Hawaii, Honolulu, HI (USA)}

\begin{document}

\maketitle

\begin{abstract}
    
We report on the upgraded One Degree Imager (ODI) at the WIYN
3.5 meter telescope at the Kitt Peak Observatory after the focal plane was
expanded by an additional seventeen detectors in spring 2015. The now thirty
Orthogonal Transfer Array CCD detectors provide a total field of view of 40’ x
48’ on the sky. The newly added detectors underwent a design revision to
mitigate reduced charge transfer efficiency under low light conditions. We
discuss the performance of the focal plane and challenges in the photometric
calibration of the wide field of view, helped by the addition of telescope
baffles. In a parallel project, we upgraded the instrument's three filter arm
mechanisms, where a degrading worm-gear mechanism was replaced by a chain drive
that is operating faster and with high reliability. Three more filters,
a u' band and two narrow band filters were added to the instrument's complement,
with two additional narrow band filters currently in procurement (including an
H$\alpha$ filter). We  review the lessons learned during nearly three years
of operating the instrument in the observatory environment and discuss
infrastructure upgrades that were driven by ODI's needs.

\end{abstract}



\section{Introduction}

The One Degree Imager (ODI) has been the major instrument development project
from 2002 to 2015 at the WIYN 3.5 meter telescope (Kitt Peak, Arizona), and its
design and progress has been documented in various SPIE conference contributions
\cite{jacoby2002, Harbeck2008, Jacoby2008, Yeatts2008, Muller2008, Harbeck2010,
    Yeatts2010, harbeck2014, gopu2014}. The instrument is designed for a one degree 
square field of view with 64 Orthogonal Transfer Array (OTA) CCD sensors. OTA
detectors\cite{burke2004, Lesser2012} were  developed to move charge in both
dimension on the detector during a science integration, allowing to actively
compensate for image motion caused by telescope tracking errors or even
atmospheric turbulence by applying corrections within an isokinetic patch size
of about 4 arc minutes on sky; this latter concept was dubbed "rubber focal
plane" by Tonry et al.~(2002) \cite{tonry2002}. ODI was designed to deliver
atmospheric limited image quality even under the best possible observing
conditions.

A first incarnation of ODI was deployed in the summer of 2012 with a partially
populated focal plane with 13 out of 64 detectors installed in a central 3
$\times$ 3 array, and four detectors placed to asses the image quality over the
field of view. This incarnation of ODI was called pODI and was described in
detail at the SPIE Astronomical Telescopes + Instrumentation conference in
2014\cite{harbeck2014}. pODI demonstrated that the instrument indeed delivered
better than 0.4" images over the entire field of view in the g', r', i', and z'
passbands.

In 2015, ODI was upgraded and redeployed at WIYN  with an enlarged 5x6  detector
array by adding 17 new detectors, providing a contiguous field of view of $40'
\times 48'$ on sky; this incarnation  of ODI is generally  referred to as "5x6
ODI", or simply ODI, as no additional development of the focal plane is planned
at this point. In this paper we report on the status of the instrument and the
upgrade process.

\section{Upgrade of the Instrument} 

ODI is designed to use Orthogonal Transfer Array (OTA) CCD detectors that have
the unique capability to move charge on both imaging dimensions during an
integration in order to compensate for unwanted image motion. This concept has
been demonstrated in pODI.  The detectors in pODI (production Lot 6) had
problems with (i) amplifier glow and (ii)  charge transfer efficiency under low
light level conditions ("fat zero" problem). The amplifier glow is mitigated by
throttling the output drain voltage of the detectors during integration. The fat
zero problem was traced to a structure between the imaging area and the serial
output register.

A  12 wafer demonstration run (Lot 7)  showed that the low light level transfer
inefficiency issue was resolved by a modification\cite{harbeck2014}, and
subsequently all wafers of that lot were processed into packaged detectors at
Imaging Technology Laboratory (ITL, Tucson, AZ). The  48 detector dies on the 12
wafers (there are 4 OTA dies on each wafer) yielded a total of 16 operational
Lot 7 OTA detectors designated for the ODI focal plane. After processing these
wafers, the processing equipment for ODI OTA detectors was retired at ITL.

\begin{figure}
	\centering
	\hfill
	\includegraphics[height=0.4\textwidth]{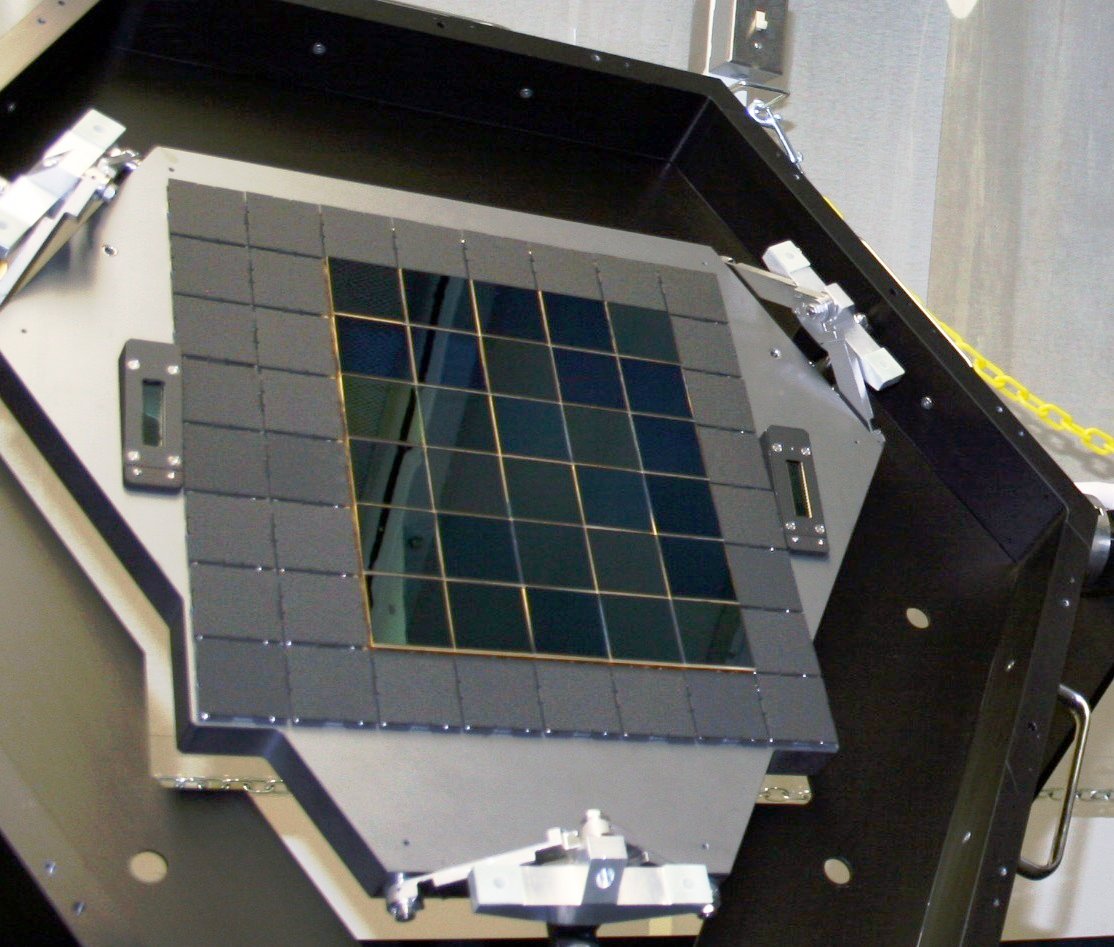}
	\hfill
	\includegraphics[height=0.4\textwidth]{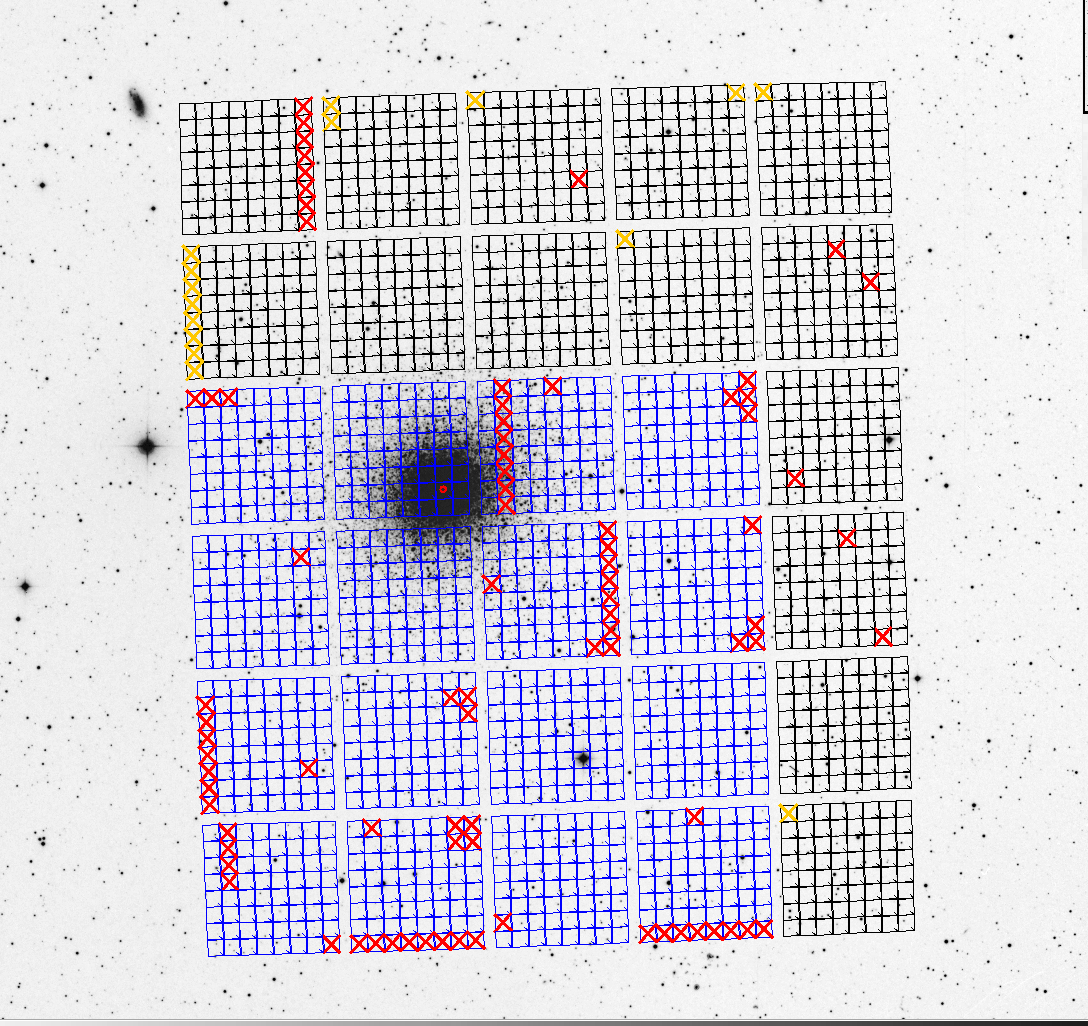}
	\hfill \\[1ex]
	
	\caption{\label{fig_focalplane} Left: Upgraded focal plane with 30 detectors
    installed in a 5 x 6 array. Each of the 30 OTAs is 50mm $\times$ 50mm in size.
    The remaining detector positions are covered with blackout plates to prevent
    stray light. Right: Imprint of the detectors on the backdrop of a globular
    cluster. The total field of view is of order of $40' \times 48'$.  Blue
    outlines indicate new Lot 7 detectors, while black outlines show the Lot 6
    units. Each OTA is comprised of 8  $\times$ 8 cells spanning about 1' $\times$
    1' on sky. Known defective cells are marked with a red cross.  Note that the
    populated focal plane is offset by half the size of an OTA detector in R.A.
    with respect to the telescope's optical axis. }
\end{figure}

The instrument was taken out of commission for upgrading from November 2014 to
May 2015. The focal plane was immediately removed from the instrument and sent
to  Imaging Technology Laboratory to add the 16 new detectors; an additional Lot
6 device was identified and added to the focal plane as well. The 14 Lot 6 and
16 Lot 7 detectors were mounted on the focal plane to result in a contiguous
central 4x4 area of Lot 7 detectors, with the remaining Lot 6 detectors filling
up the surrounding area to produce a 5 by 6 detector array (See Figure
\ref{fig_focalplane}).

While the focal plane was repopulated with detectors, the cryostat and the 
instrument hardware were refurbished. We found in particular:
\begin{itemize}
	\item Some epoxy-sealed vacuum feed throughs had developed  leaks. We 
	attribute those to the thermal cycling caused by the nearby CCD controllers. 
	The daytime idle position of the instrument was changed to prevent the 
	formation of heat pockets near the vacuum seals, and to allow a better
	convective cooling flow in the CCD controller chassis.
	
	\item Inside the dewar we have identified areas of outgassing residue, which 
	were cleaned.

	\item The optical surfaces of ODI's corrector optics needed cleaning,
	 including the inside of the cryostat's dewar window. 
	
	\item The lubricant in a ring bearing for the atmospheric dispersion corrector
(ADC) prism had separated, and the liquid oil component started to flow over
one of the prisms. The optics were cleaned with no detrimental effect to the
surface and its anti-reflection coating. Since the ADC bearings operate only at
a very slow speed, the lubricant was entirely removed from the ring bearings to
avoid the risk of future leaks.

\item  At the time of the dewar refurbishment, the molecular sieve material in
the dewar (Zeolite 5-A) saturated (presumably with water), leading to an
unmanageably short vacuum hold time of a few days, and it was hence replaced
with fresh 5-A material. In the past, the Zeolite 5-A had been replaced on an
annual basis due to its saturation. Since replacing the Zeolite in the dewar is
an expensive and labour intensive operation, we started to investigate
hydrophobic Zeolite ZSM-5 as a replacement candidate. Since ZSM-5 does not
permanently bind water as 5-A does, it can be regenerated by vacuum pumping at
room temperature alone, without the need for high temperature backing.

In January 2017 the ODI molecular sieve material was replaced with Zeolite
ZSM-5. After over a year in operation, the ZSM-5 material still performs well
and shows no signs of degradation. We note that the ODI dewar is warmed up and
vacuum-pumped on a regular basis (every few months), which regenerates the
ZSM-5's ability to adsorb gas.

\end{itemize}

\subsection{Focal Plane Performance}

As expected, the Lot 7 OTA detectors performed similarly to the Lot 6 detectors,
but without displaying the low light level charge transfer efficiency problem.
Since the 5x6 ODI focal plane utilizes both Lot 6 and Lot 7 detectors,
observations that depend on the full field of view still require a minimum
background level of the order of 100 electrons, which sets a lower limit on the
useful exposure time. If the smaller field of view of the continuous Lot 7
sub-array is sufficient for the scientific goals of an observation, those
restrictions do not apply. The amplifier glow in the ODI OTA detectors remains
unchanged, and, as in pODI, it is mitigated by throttling the output drain (OD)
voltage while idling or integrating light,  and only turning it fully on during
the readout of a detector. Detectors that are used for guiding during an
exposure have their OD turned fully on. This  contaminates large parts of
that detector with amplifier glow, making it unusable for science for that
exposure.

Changing the OD of the OTA detectors during integration and idling versus
reading out leads to a change in their power dissipation. We estimated the
difference between the OD throttled and fully on to be about 0.5W per detector.
Reading the detectors while OD is fully on will hence dissipate an additional 15
Watts into the focal plane for the read time of about 6 seconds. This causes
temperature variations in the focal plane of the order of several  tenths of to
a full degree Kelvin, in particular during high cadence use cases. While the
total cooling power of the ODI focal plane would be sufficient to cool a focal
plane that was fully populated with 64 OTA detectors, the accuracy of the
temperature control might not meet the specifications with a full compliment of
OTA detectors.

The Stargrasp CCD controller\cite{onaka2008} used in ODI can control one or two
detectors per board. For pODI, a maximum of one detector was connected per
controller board, and the cross-talk behavior for two connected detectors was
not yet demonstrated in pODI. In the 5x6 ODI focal plane, 2/3 of the detectors
share a controller board, whereas 1/3 of the detectors are controlled by a
single Stargrasp board each. No evidence of cross-talk between the two channels
of a Stargrasp board was found.

The detectors in the pODI instrument were placed to sparsely sample the entire
1x1 degree field of view of ODI, and the delivered image quality of pODI was
found to be better than 0.4" on sky over the entire field of view in the ODI
g',r',i', and z' band filters. The same performance was confirmed for the 5x6
ODI focal plane, i.e., the alignment was reproduced when the focal plane was
removed and reinstalled into the dewar for the detector upgrade.

Further details about the instrument performance are described in the ODI User's
Manual\footnote{\url{http://www.wiyn.org/ODI}}.

\subsection{Tuning the acquisition software}

The ODI data acquisition software is described in detail in previous SPIE
contributions\cite{Yeatts2008,Yeatts2010}. In short, the 5x6 ODI focal plane is
read out by 20 Stargrasp CCD controllers which interface to a network switch via
individual  1GB ethernet over fiber connections. The switch itself bundles the
data connection into a 10GB network backbone used by the ODI computers (three
units with 24 cores, 32GB RAM each). The data stream from the 30 CCDS is
collected by a single server / Java virtual machine that is managed by a JBOSS 5
application server. From there, data are saved to a storage device.

The data rate has more than doubled (from 13 detectors to now 30) to about 1
Gigabyte that downloads from the detectors in about 6 seconds.  The scalability
of the data acquisition system was tested before the upgrade by creating a
virtual focal plane  configuration to simulate the enlarged focal plane. This
test revealed some bottlenecks: the interval between sustained bias readouts
increased from about 25 seconds to over 40 seconds. Several bottlenecks in the
data acquisition code were identified and mitigated during the upgrade project
and commissioning phase:

\begin{enumerate} 
    
\item For pODI, all data were saved directly to an NFS-mounted storage devices (Oracle ZFS
appliance) and the delay in  writing FITS files to disk was the major factor
slowing the data handling process. As a mitigation we equipped the acquisition
server with a Raid 5 configured local storage array, utilizing only disks with a
6G interface; new images are now first stored on the local hard drives. The
image data are then slowly transferred by a background process to the storage
appliance, which serves both as a mid-term storage and to stage data for
transfer into the ODI data archive.

\item   Simultaneously receiving the data streams from 30 detectors via
TCP/IP posed no major challenge for the 10GB network backbone.  However, when
writing the data to disk we realized a significant  improvement in the write
speed by limiting the number of data streams that are written in parallel to
disk. As a mitigation, the FITS file writing routine was changed into a Java
{\tt Callable} that would be submitted to a multi-threaded Executor, and we
determined that ten storage threads delivered a good performance. The write
performance was further improved by rewriting the {\tt nom.tam.fits} package to
allow writing into {\tt BufferedOutputfile}.

\item As part of the data storage process, a thumbnail image is generated for
each detector to help the observer to judge image quality in real time. For
pODI, the thumbnail generation was delegated to a command line utility that was
called after the images were written to disk, reading back (and decoding) the
FITS file again. Also, the thumbnail generation process was unnecessarily
serialized in the data acquisition process, i.e., telescope observations would
be blocked until thumbnail images were generated.  For 5x6 ODI we have moved the
thumbnail generation into the Java virtual machine, where the data are already in
memory, hence saving I/O and CPU time for FITS decoding. Furthermore, the
generation of the thumbnails was delegated to  lower priority background
threads, making the thumbnail generation an asynchronous process.

\end{enumerate}

As indicated by prior testing, the existing IT infrastructure was capable of
handling the larger focal plane after some investments into faster local storage
and by serializing the data storage  processes.  After each readout, about 6
seconds are spent to flush the detector to remove residual charge, and the
readout overhead is now limited by the detector and telescope performance.

\subsection{Advanced OTA modes}

Over the time of the ODI project we have demonstrated that OTA detectors can
successfully compensate for tip / tilt image motion at an approximate 20Hz rate
by moving charge on the detector; this ability was the original motivation to
utilize OTA detectors in ODI.  However, the amplifier glow in ODI's OTAs
prohibits the use of a detector for simultaneous guide star acquisition and
science integration; localized atmospheric image motion compensation within an
isokinetic patch (about 4 arcminutes on sky, or a quarter section of a single
OTA detector) is hence not practical. Even if the amplifier glow were to be
addressed successfully, there are practical limitations to implementing a rubber
focal plane as envisioned in Tonry et al.~(2002)\cite{tonry2002}. The density of 
bright guide stars is insufficient in some  optical passbands, in particular
outside the galactic plane. It is not sufficiently established how one would
proceed with the astrometric calibration of such a rubber focal plane, and the
data processing effort would be significant for a modest improvement in
delivered image quality. Neither the PanSTARRS survey nor the ODI instrument
have advanced the rubber mode into turn-key operations.

The on-chip image motion compensation for ODI can still be useful to correct for
telescope tracking errors or light wind shake by sensing the average motion of
several guide stars, e.g., at the corners of the field of view; averaging their
signal would ignore localized atmospheric turbulence. We call this the "coherent
guide mode"\cite{harbeck2014}. This observing mode is implemented in the ODI
software and theoretically usable by observers on demand, but at this time there
is no support in the  data processing pipeline. Also, ideally one would use
three to four bright guide stars in the corners of the field of view; since the
use of a detector for guide star acquisition requires the output drain to be
turned on during the exposure, amplifier glow will render those detectors
unsuitable for science use, reducing the usable science field of view. To date,
the coherent guide mode has not gained traction beyond the technical
demonstration.

\section{Filter change mechanism upgrade and performance}

\begin{figure}
	
	\hfill
	\includegraphics[height=0.49\textwidth]{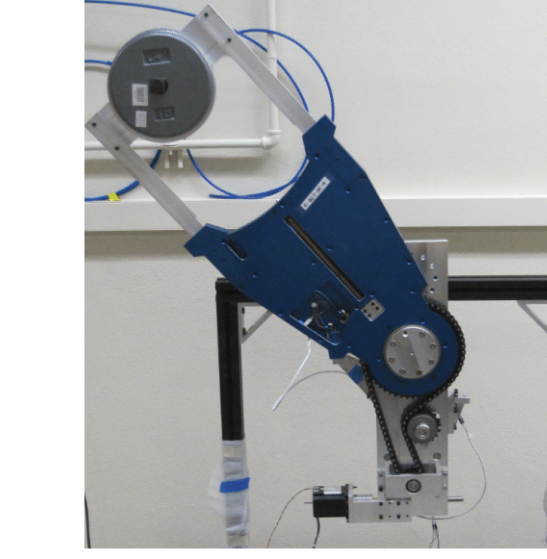}	\hfill
	\includegraphics[height=0.49\textwidth]{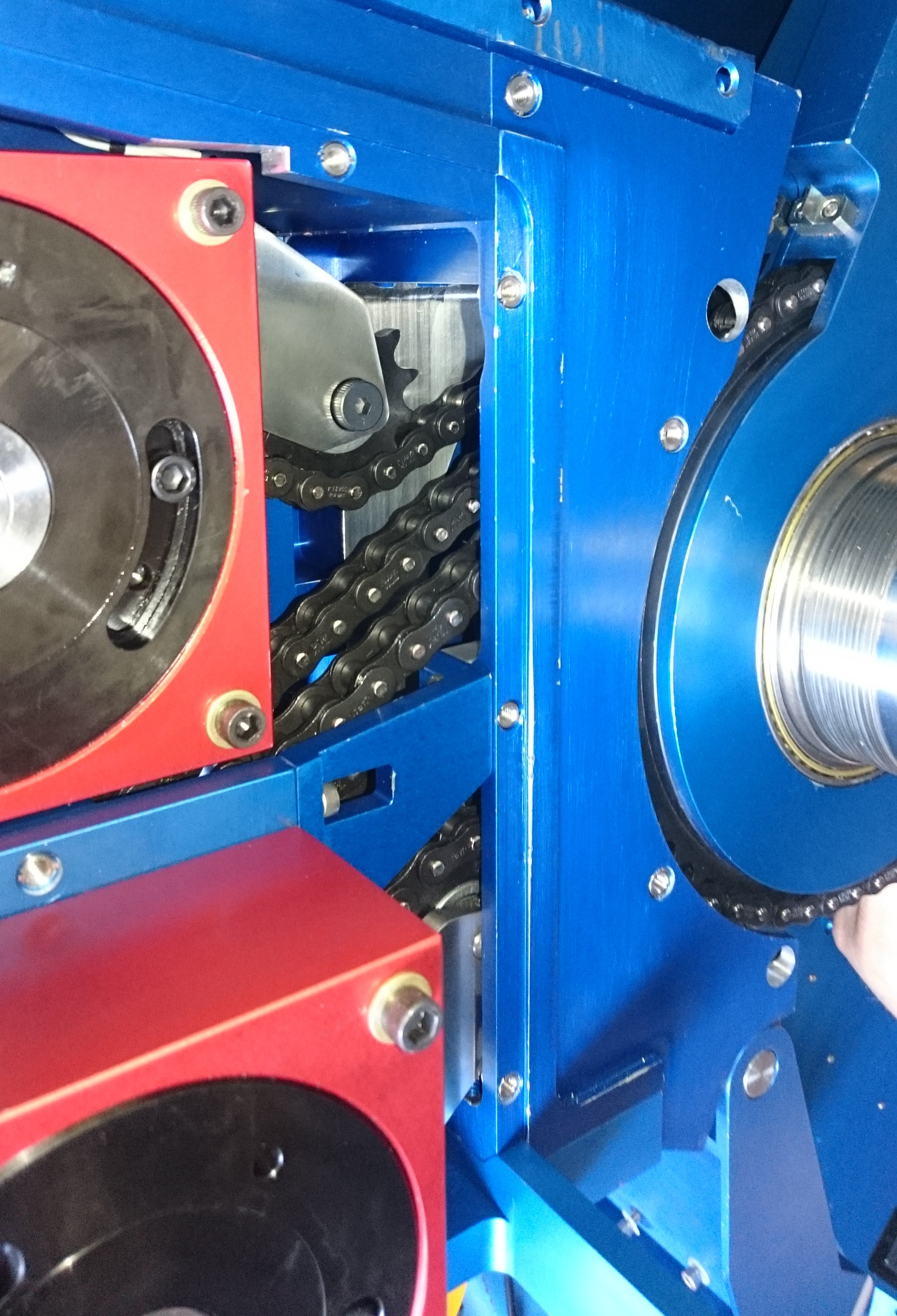} \hfill \hfill
    
	\caption{\label{fig_filterdrive} Left: Single filter arm test assembly. The
    42cm $\times$ 42cm filter is simulated by a weight. This assembly was used to stress-test
    the filter arm drive concept with order of 6000 in and out actuations.
    Right: Close-up of the three-filter layer assembly during installation at the
    instrument.} 

\end{figure}

ODI has a filter change mechanism that can hold up to nine filters, where each
filter is about $42 \rm{cm} \times 42 \rm{cm}$ in size. Each filter is inserted
and removed out of the beam by a semaphore-like filter arm that was originally
driven by a worm gear\cite{Muller2008}. Due to the proximity to optical
surfaces, the worm gear contact was operating without lubrication (steel on
bronze contact). During integration testing of pODI in 2012 it was realized that
this drive would not be viable for long-term operations, as the stainless steel
worm gear would slowly shave off the bronze gear on the filter arm. However, due
to time and resource constraints, pODI was deployed as it was with the
understanding that this issue would have to be addressed eventually. In the
meantime, the bronze gear was replaced on an annual basis due to wear and tear,
but even with fresh gears operation was unreliable because of increased
friction, and occasionally bronze shavings caused short circuits in the filter
arm position sensors.

In 2014 we  started a project to redesign the filter drive, where instead of a
worm gear the filter arm would be driven directly by a chain drive. The original
custom-made gear box of the filter drive was replaced by a commercial,
encapsulated lubricated worm gear. A single chain-driven filter arm prototype
was designed and built at the University of Wisconsin-Madison (Figure
\ref{fig_filterdrive}, left), and its operation was successfully demonstrated by
over in/out 6000 actuations  under several spatial orientations. After
testing, the worm gear box was taken apart and inspected for signs of excessive
wear - none  could be identified.

Upon successful testing, a full complement of nine filter drives (three units
that serve three filter arms each) was built at
the NOAO instrument shop, and the upgraded filter drive was installed into the
ODI instrument during the summer of 2015 (see Figure \ref{fig_filterdrive},
right). The new drive has worked without problems ever since.

\section{Stray light and mitigation by additional baffles}

\begin{figure}
\centering 
\includegraphics[height=0.28\textwidth]{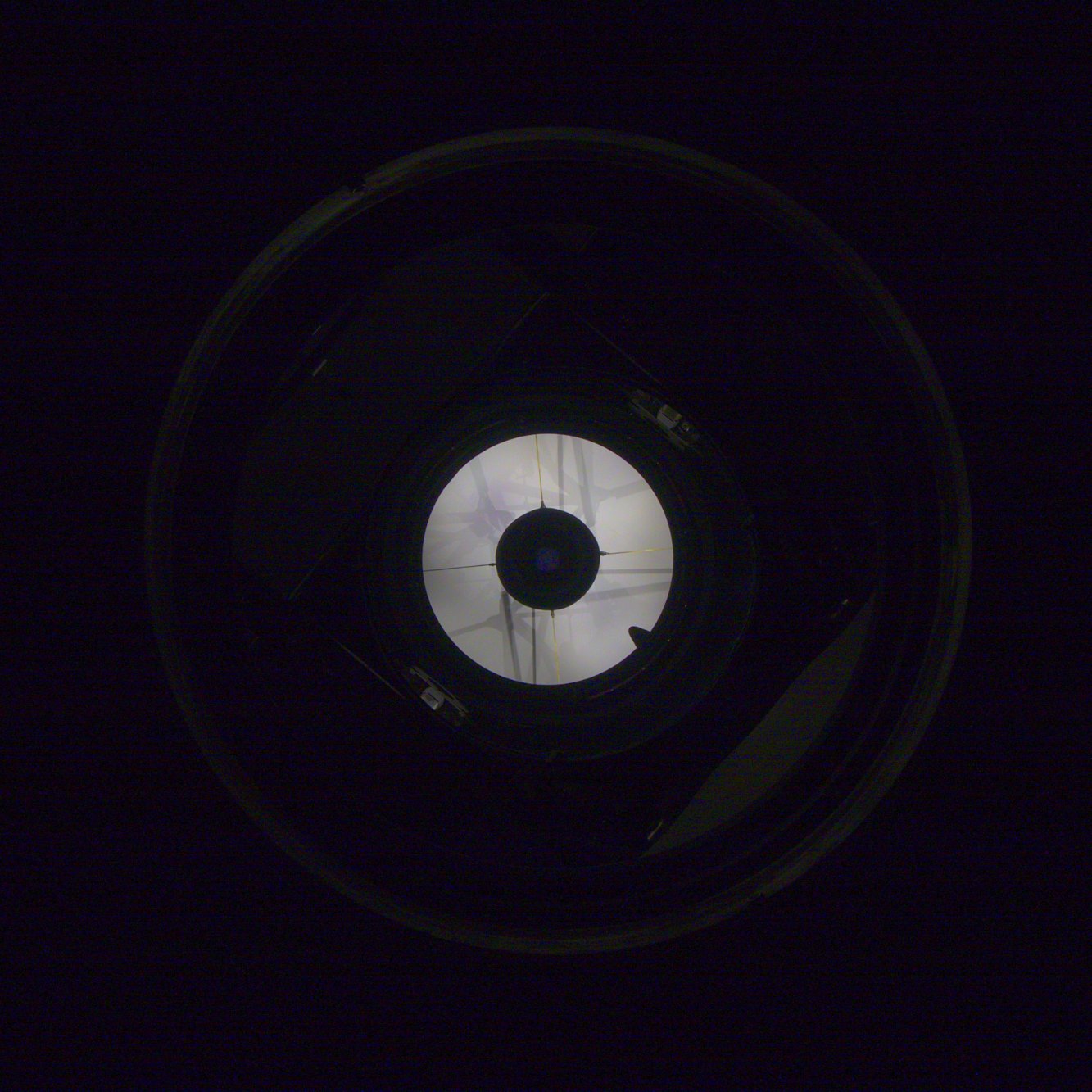}
\includegraphics[height=0.28\textwidth]{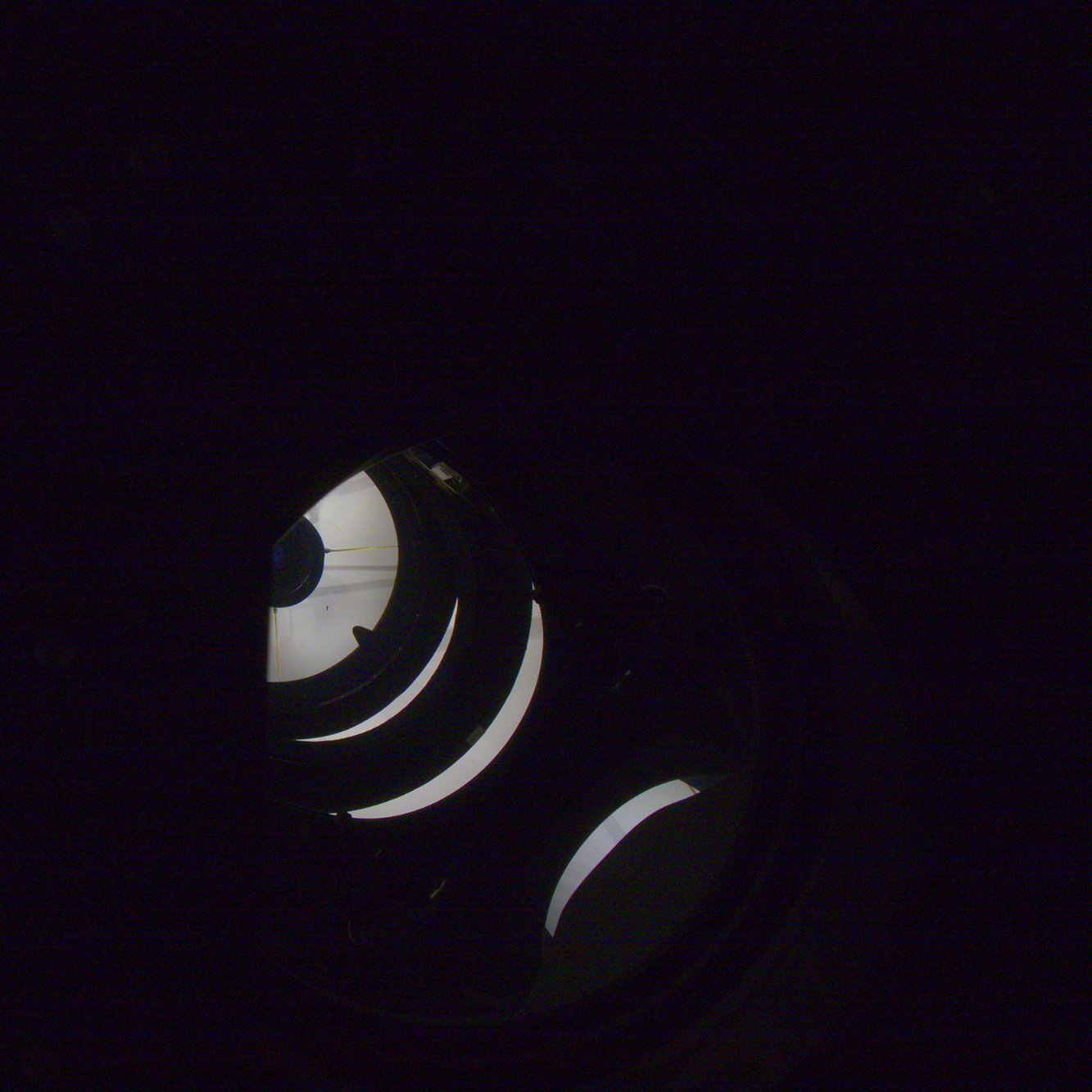}
\includegraphics[height=0.28\textwidth]{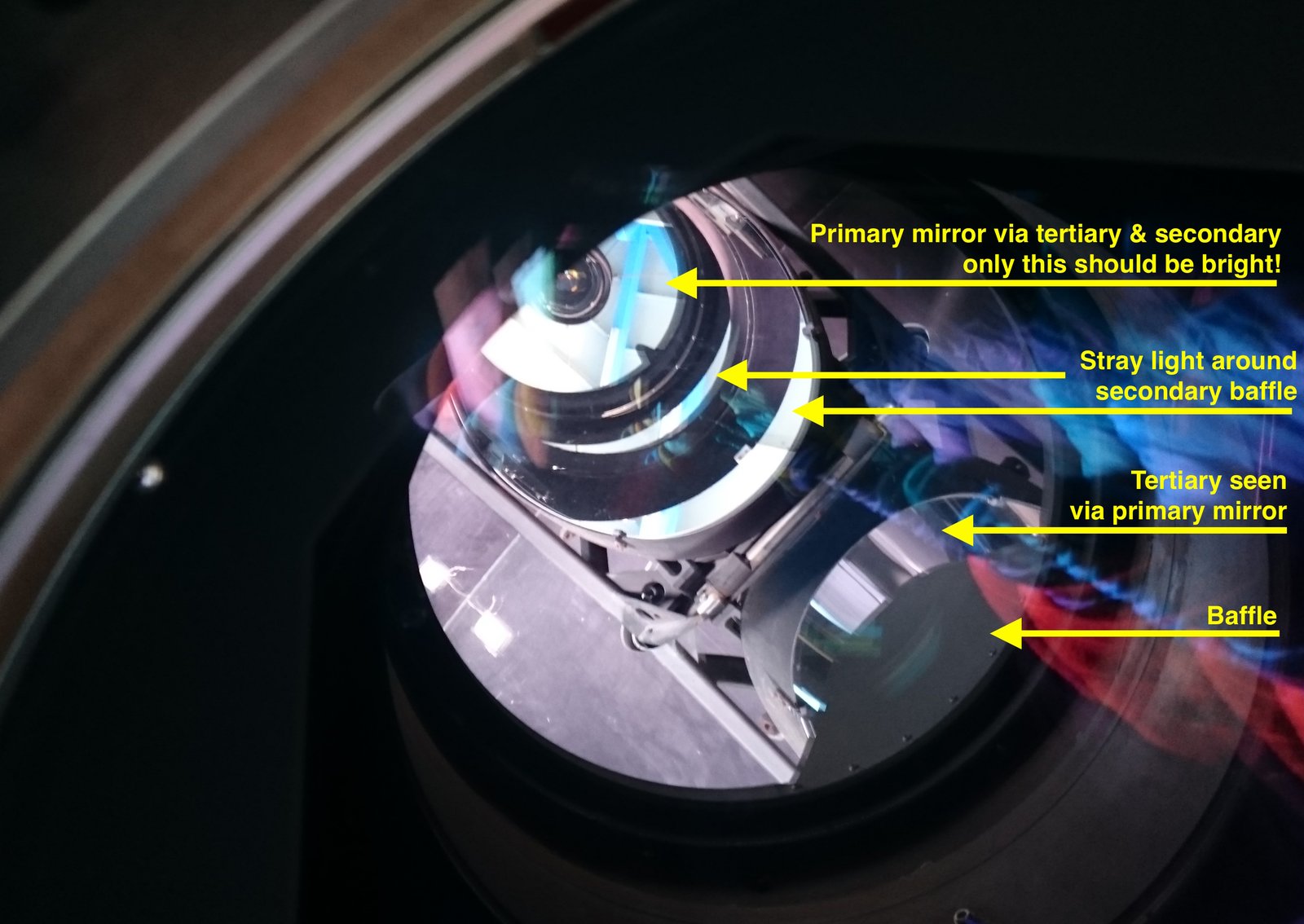} \\[1ex]

\caption{\label{fig_straylight} View through the ODI instrument and telescope
    onto the WIYN flat field screen. Left: at the center of the field, where only
    the mirror pupil is a major contributor of illumination. Middle: View from the
    upper left corner of the instrument. Additional stray light arcs become visible.
    The two central  arcs originate from the off-field light that enters the instrument
    directly from a reflection off the tertiary mirror. The  lower right arc is
    due to reflection of off field light off the tertiary, onto the primary mirror,
    and  into the instrument. Right: Same as the middle, but with ambient lights 
    switched on. The individual stray light contributors are identified.  }
\end{figure}

In advance of pODI's original  deployment, a detailed stray light analysis
executed by Photon Engineering (Tucson, AZ) indicated that a significant
contribution of stray light was to be expected on the ODI focal plane. The two
major contributors are (i) off-axis rays reflecting off the tertiary mirror,
then the primary mirror, finally entering the instrument and (ii) off-axis rays
reflecting directly off the tertiary mirror, entering the instrument. The first
stray light mode had already been addressed by installing an additional baffle
at the entrance aperture of pODI. The latter stray light path, however, had not
been mitigated yet by the time of the 5x6 ODI upgrade. The origin of the stray
light is clearly visible in Figure \ref{fig_straylight}, where we show the view
through the instrument (with the dewar removed) at the center and upper left
corner of the field of view.

For the initial pODI deployment with a smaller focal plane the stray light did
not significantly affect operations since the affected area was barely covered
by detectors; the extended focal plane of 5x6 ODI however was strongly affected
by the second stray light path, where the stray light contribution to the
background is of order of 30\%. For night time observations the stray light
manifests in additional background structure that can be subtracted; for the
acquisition of flat field calibration images the stray light is detrimental.

	\begin{sidewaysfigure} 
		\centering

		\begin{tabular}{ccccc} odi u' & odi g' & odi r' & od i'
			 & odi z' \\ \hline \multicolumn{5}{c}{ratio of $\pm 90^\circ$ 
				flat
				fields, no baffle} \\[1ex]
			
			\includegraphics[width=0.18\columnwidth]{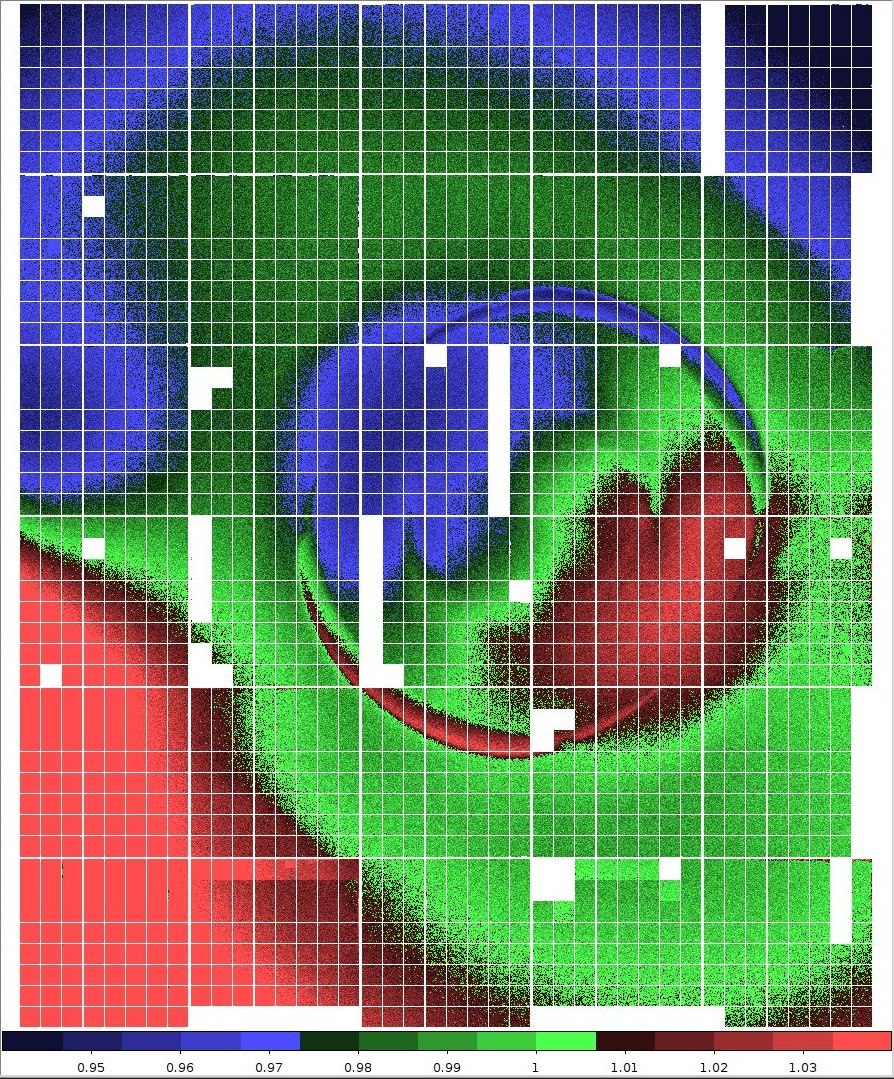}
			&
			\includegraphics[width=0.18\columnwidth]{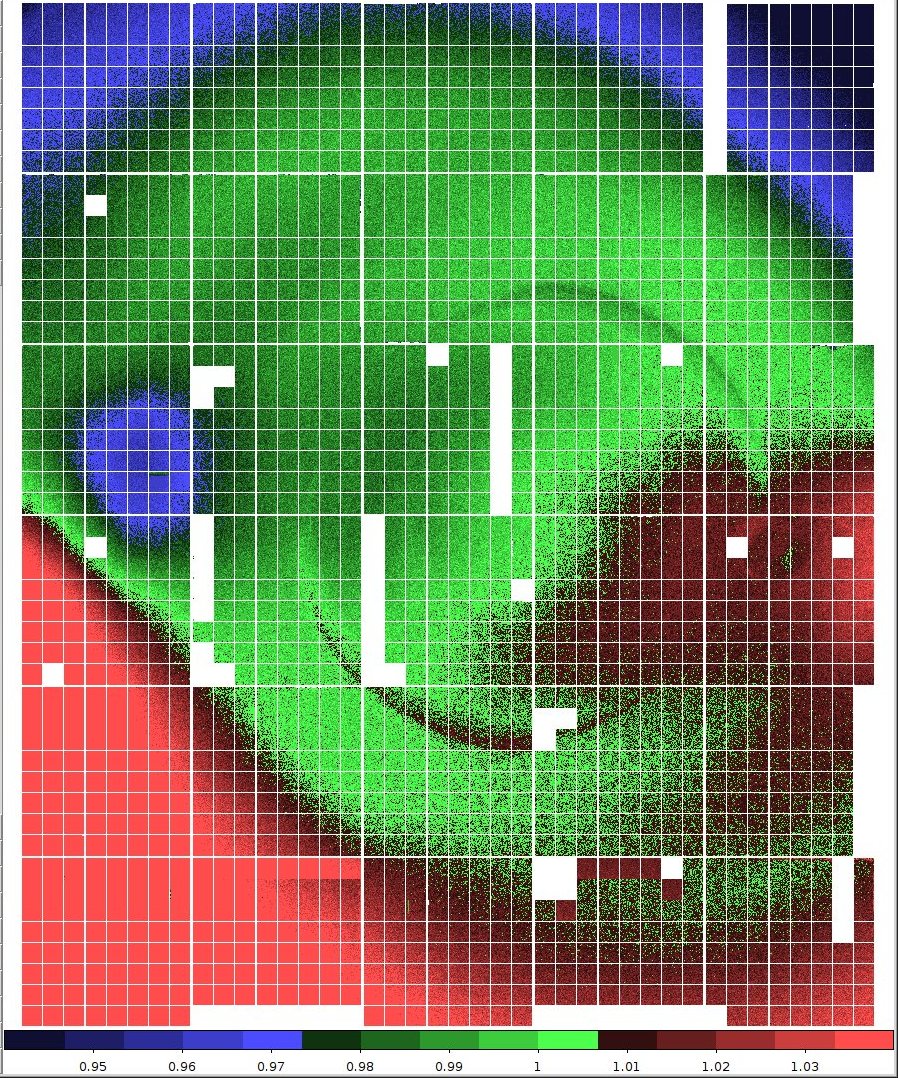}
			&
			\includegraphics[width=0.18\columnwidth]{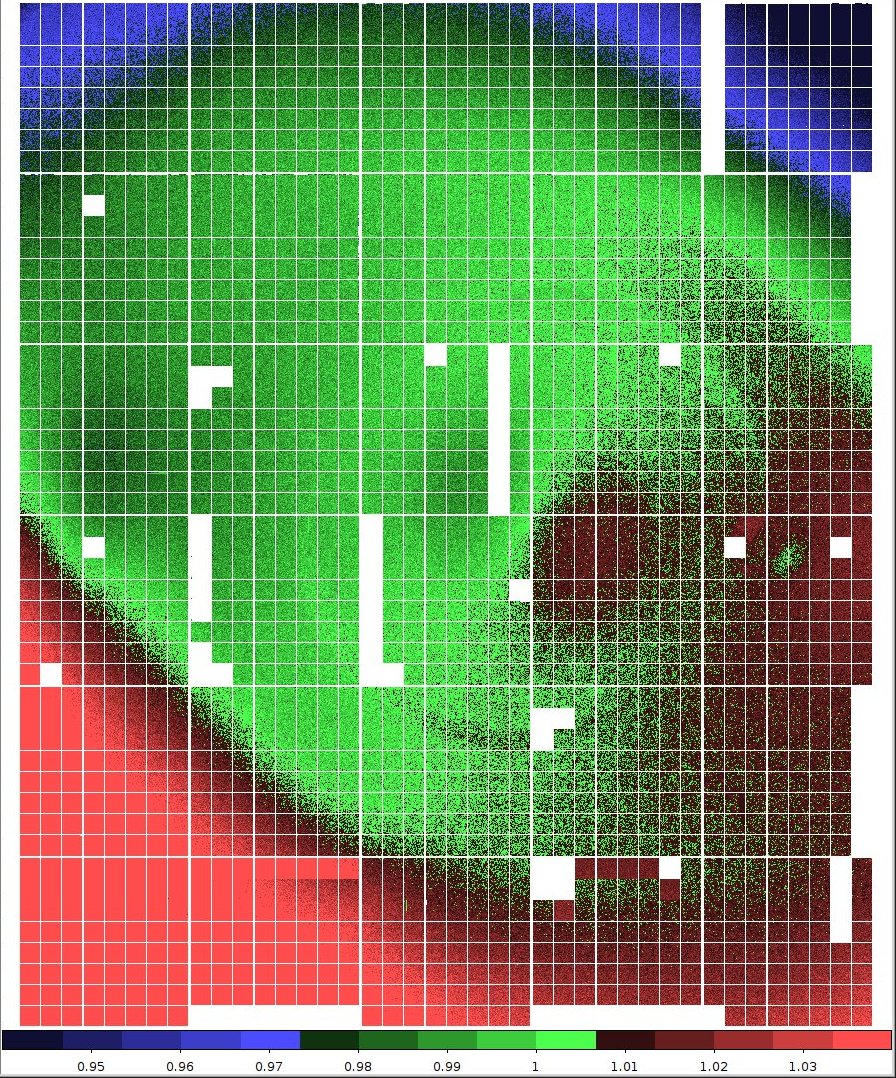}
			&
			\includegraphics[width=0.18\columnwidth]{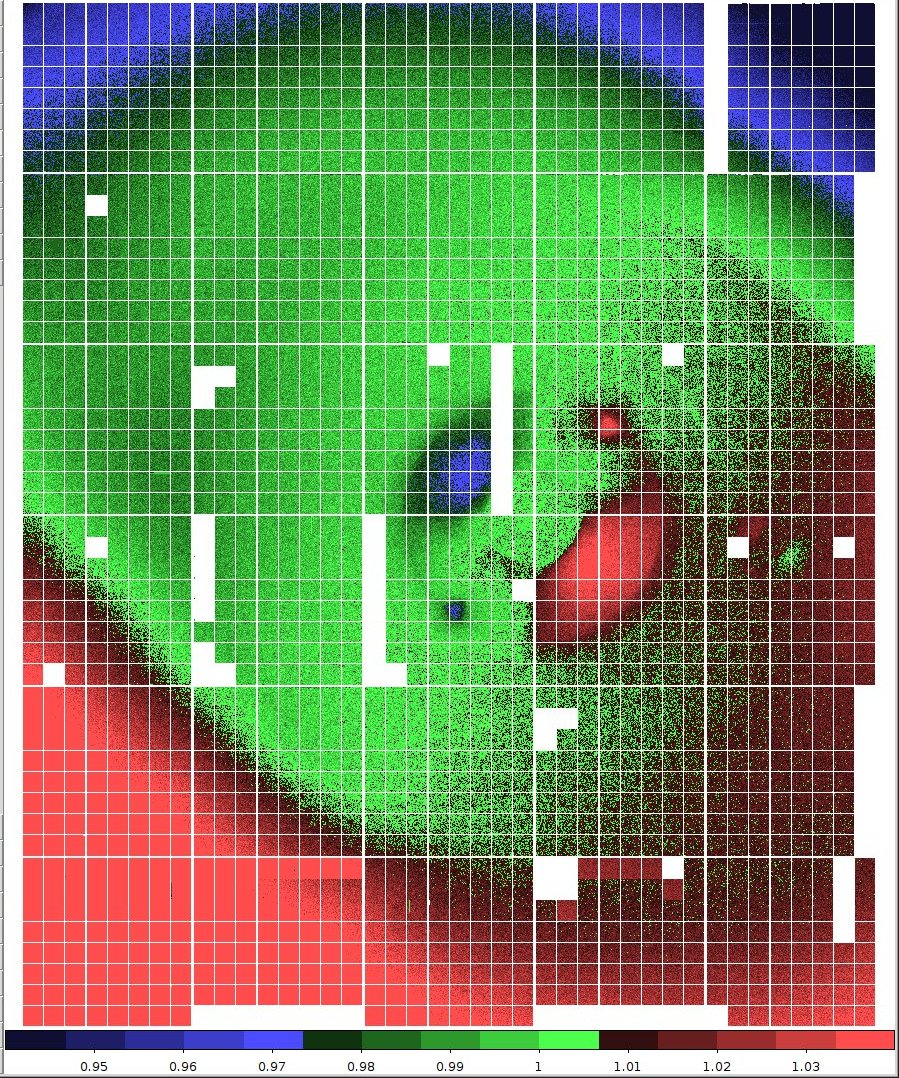}
			&
			\includegraphics[width=0.18\columnwidth]{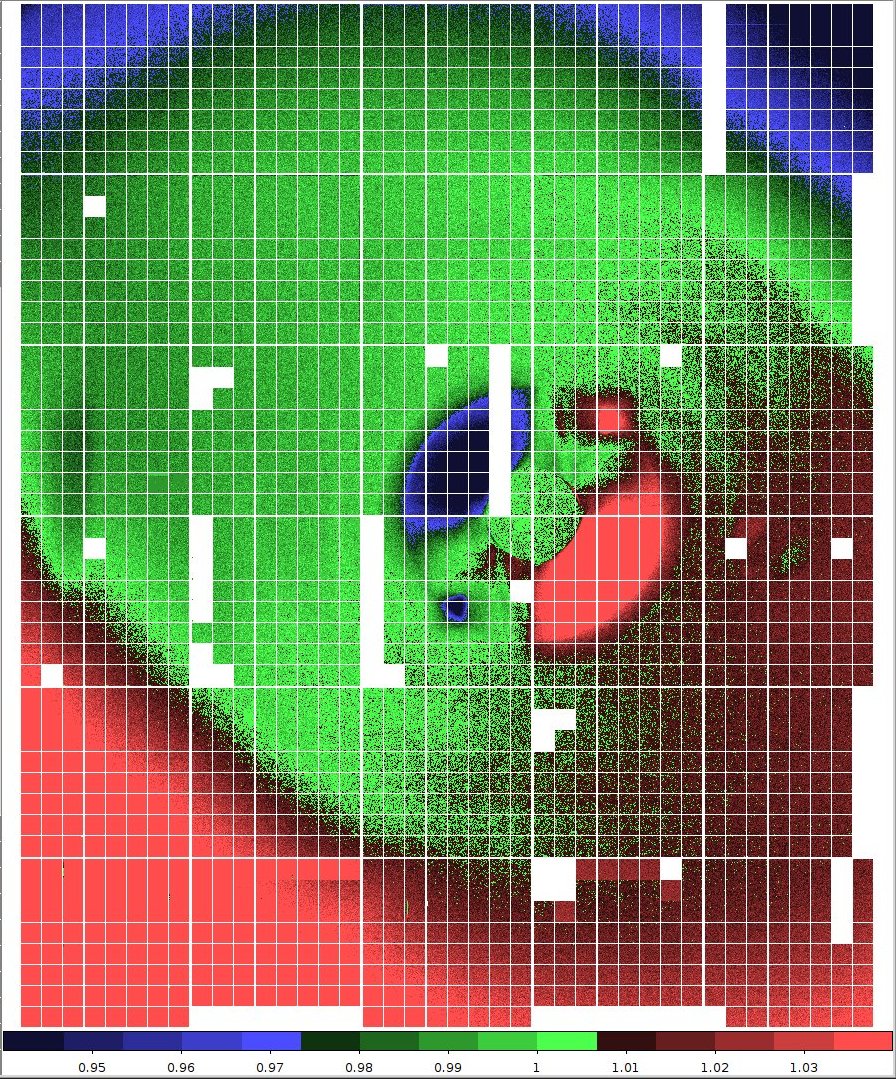}
			\\[2ex]

			\includegraphics[width=0.18\columnwidth]{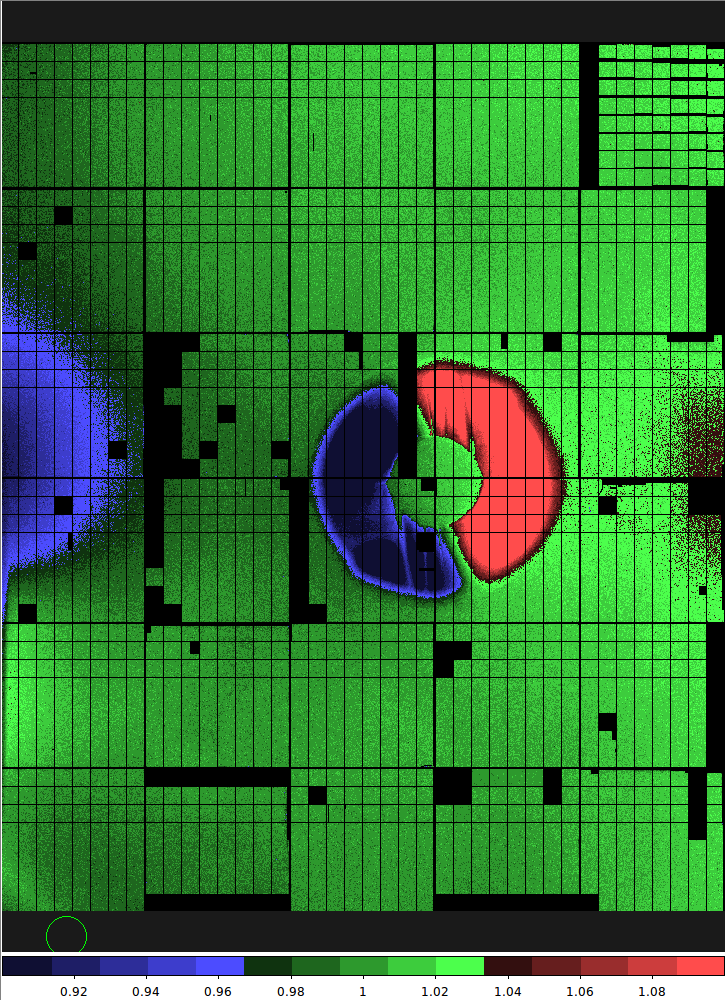} &
			\includegraphics[width=0.18\columnwidth]{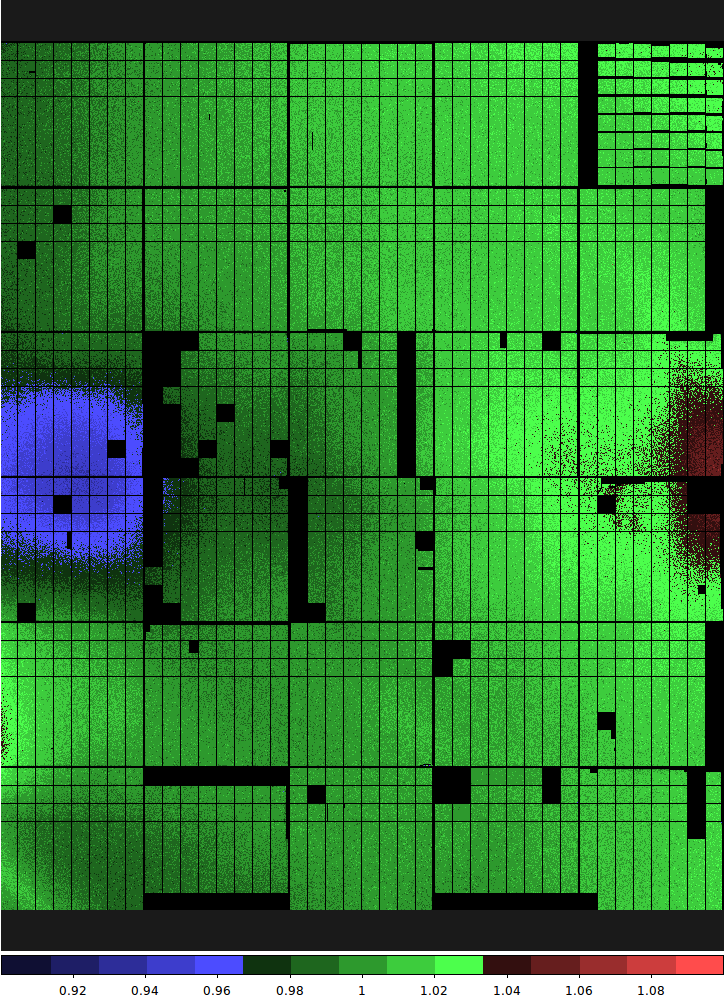} &
			\includegraphics[width=0.18\columnwidth]{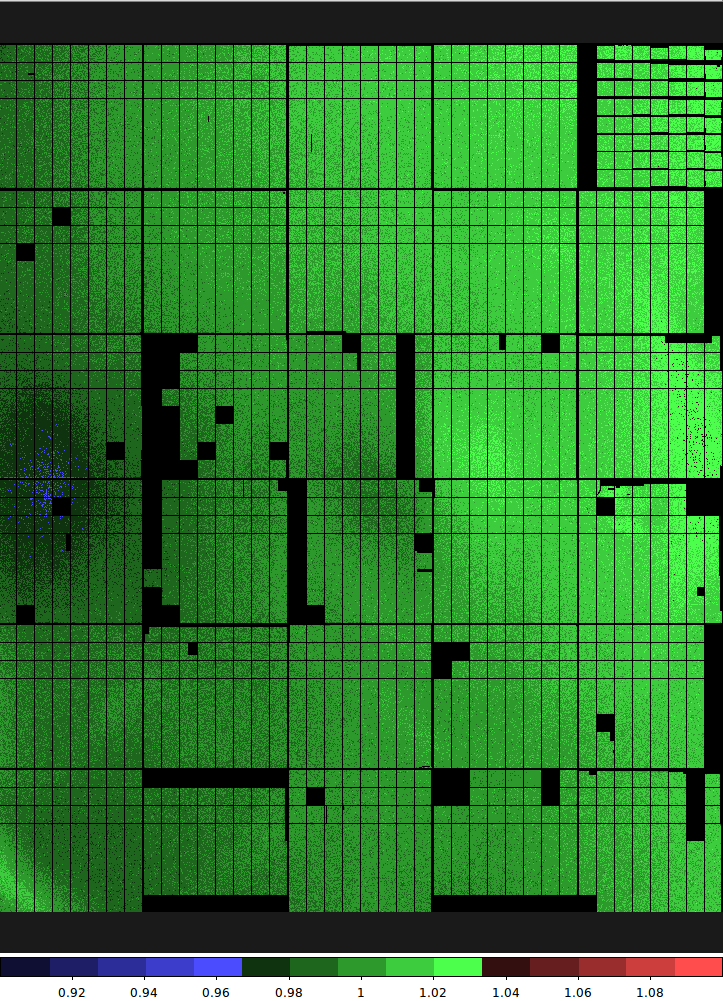} &
			\includegraphics[width=0.18\columnwidth]{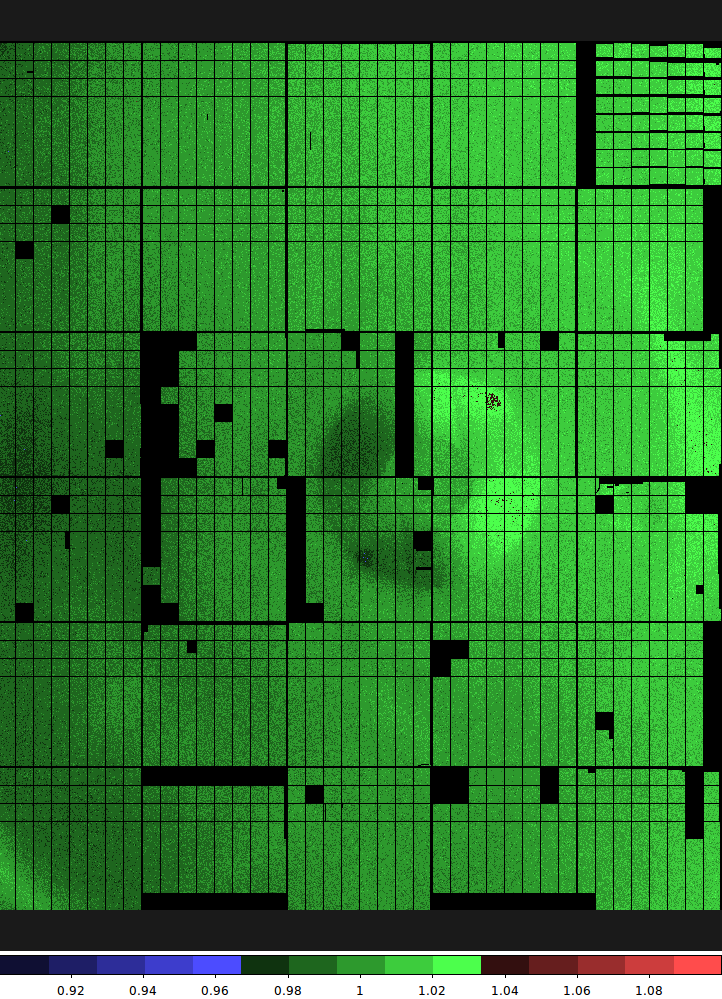} &
			\includegraphics[width=0.18\columnwidth]{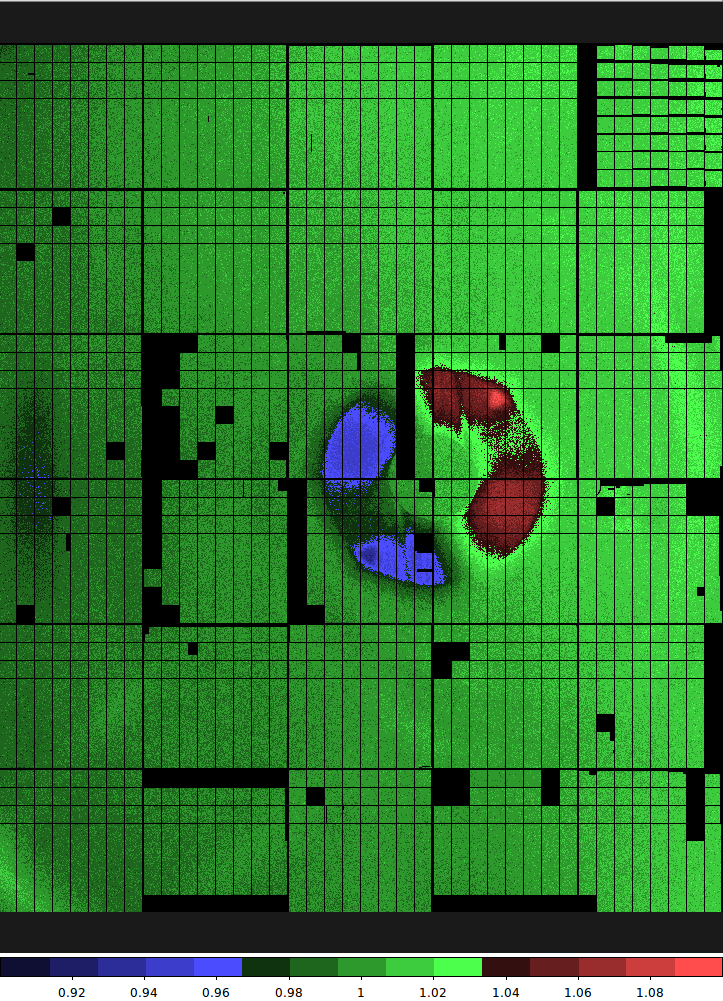} 
			\\[1ex] 
			\multicolumn{5}{c}{ratio of $\pm 90^\circ$ flat
				fields, with baffle} \\[2ex] 
		\end{tabular}
		
		\caption{\label{fig_flatfieldbaffle} Ratios of dome flat field 
			images
			taken at an instrument rotator angle of $+90^\circ$ and 
			$-90^\circ$ 
			for all ODI broad band filters. In a stray light free  
			instrument, the ratio should 
			be uniform. Top: flat field ratios before the telescope baffle 
			installation. 
			Bottom: flat field ratios after telescope baffle installation. 
			With the telescope 
			baffles installed, only ghosting from the instrument remains in 
			the flat field 
			ratio; the mitigation of this ghosting is discussed in section 
			\ref{sect_ghost}. }
		
	\end{sidewaysfigure}

The stray light  can be readily identified by comparing flat fields that are
taken at different instrument rotations. We demonstrate this in Figure
\ref{fig_flatfieldbaffle}, where for a set of the g'r'i'z' filters we show the
ratio of dome flats that were taken with a nominal instrument rotation of plus
and minus 90 degrees.  In case of a flat illumination, one would expect a
constant ratio of unity in those images. The flat field ratios however show a
strong sign of (i) central pupil ghosts (see next section) and (ii) stray light
in the lower left corner.

The stray light was mitigated by installing two additional baffles  at the telescope
according to Photon Engineering's recommendation: one larger ring baffle
around the tertiary mirror, and one slim ring around the secondary baffle. These
baffles sufficiently block the stray light paths that can be seen in Figure
\ref{fig_straylight}. The new baffles increase the central obscuration of the
telescope, causing a 2.3\% loss in collected light. For this reason, the baffles
were designed to allow for quick removal when using narrow field instruments
that are not adversely affected by the stray light. While not studied, we expect
that the Hydra fiber fed spectrograph located on the opposite Nasmyth port at
the WIYN telescope will benefit from the baffles as well.

\section{Pupil ghost suppression using a two bladed shutter}
\label{sect_ghost}

\begin{figure}
\centering
\includegraphics[height=6.5cm]{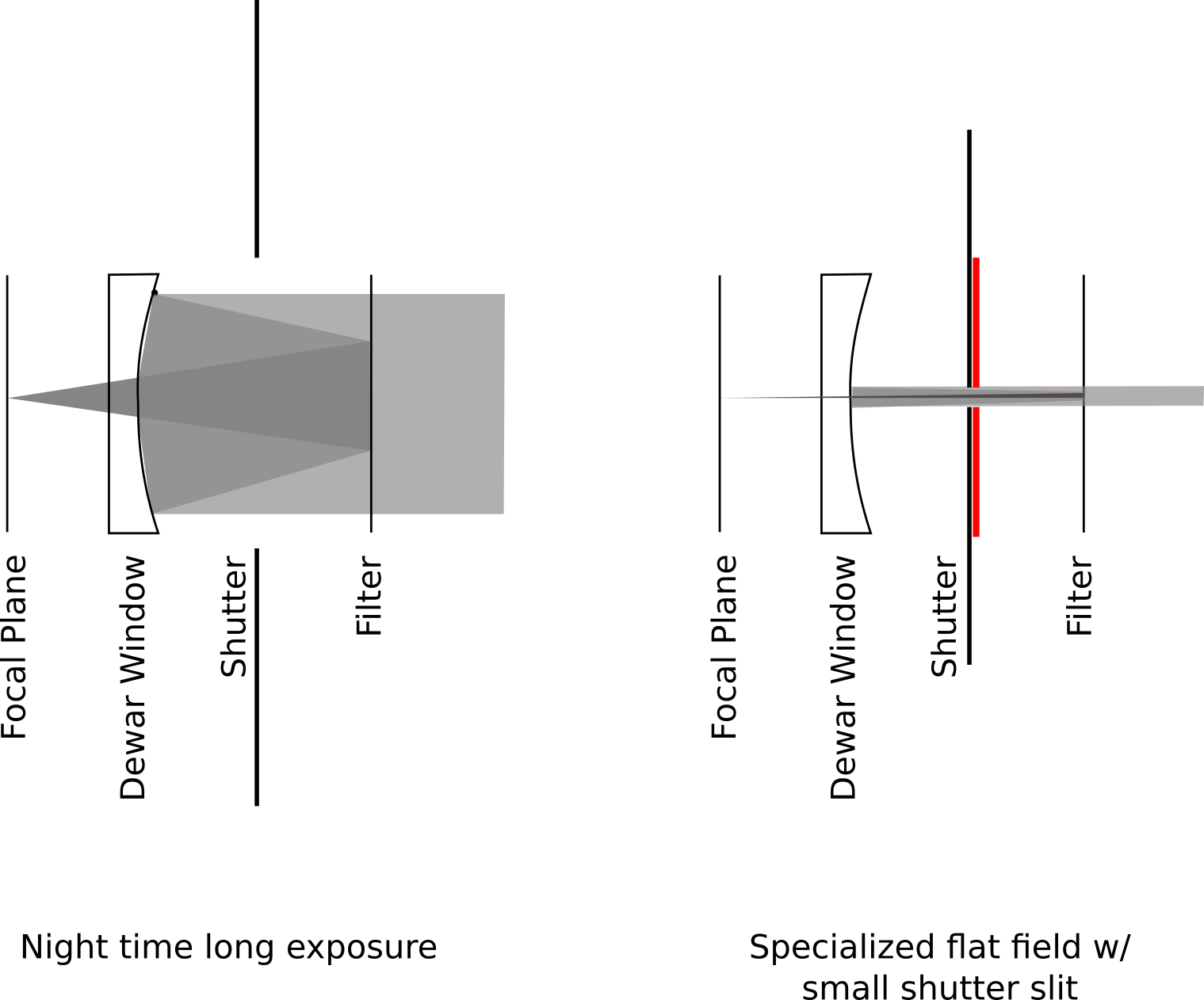}
\hspace{0.75cm} \includegraphics[height=6.5cm]{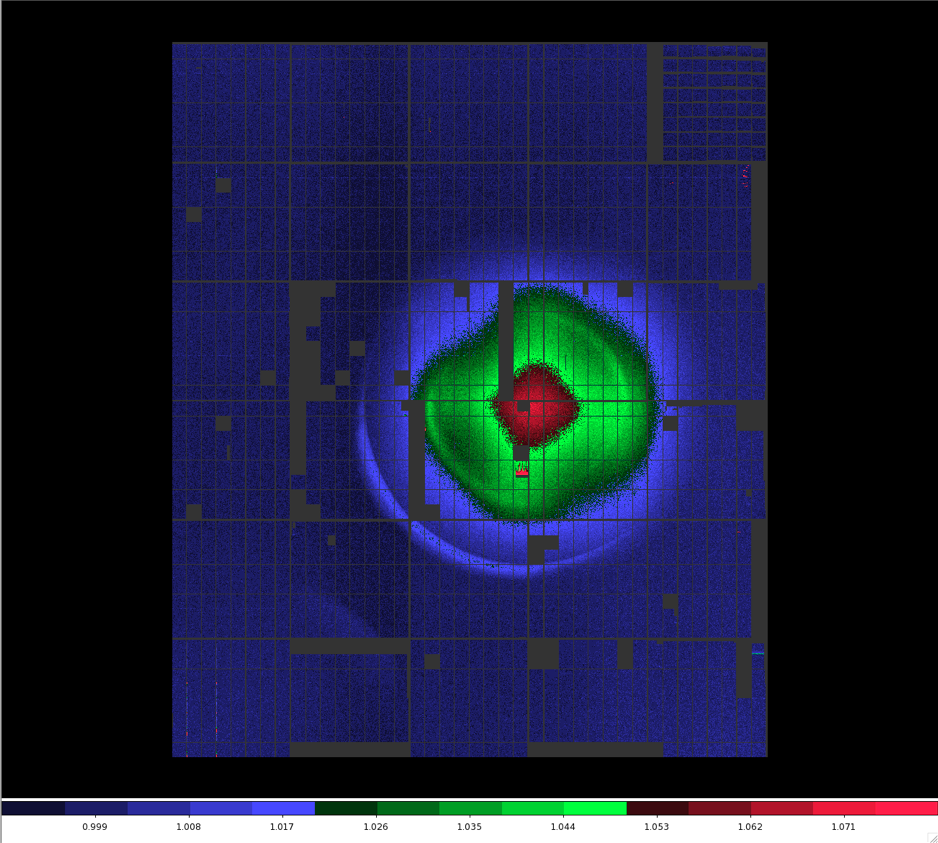}
	
\caption{ \label{fig_pupilghost}Left: Formation of the pupil ghost in ODI: 
    light entering the instrument from the telescope (from the right) reflects off
    the concave dewar window, and the converging beam reflects off the filter to form
    an in-focus image of the telescope pupil. Right: Ratio of a flat field taken
    with different shutter modes. The sensitivity variations between pixels and
    detectors cancel out (shown in dark blue, indicating that the ratio is $\approx 1$),
    but the excess ghosting light remains.}
\end{figure}

Reflections of the optical surfaces within ODI produce internal ghosting that
contaminates flat field calibration and science exposures. The major
contributing light path is a reflection of ODI’s concave dewar window, that is
then back reflected at a flat filter to form an image of the telescope pupil on
the focal plane (Figure \ref{fig_pupilghost}).  Thanks to anti-reflection
coatings on the optics, each reflection is suppressed to a 1\%-2\% level,
depending on the actual wavelength of the light, but the residual light is still
capable of producing a significant ghosting component. Due to the
wavelength-dependent performance of the anti-reflection coating, the pupil ghost
is more pronounced at shorter wavelengths. Also, since the pupil ghost forms by
reflection of a band pass filter, and the ODI filters are arranged in three
layers, the size and hence concentration of the pupil ghost will change with the
in-beam filter.

The pupil ghost is undesirable for several reasons: In night-sky observations it
will produce an extra background component, albeit one that can be mitigated by
subtracting out a model. More severely, the pupil ghost also produces an extra
light component in a calibrating flat field, and if not corrected, would lead to
a false sensitivity calibration of the affected areas.

The first approach to removing the pupil ghost from calibration images was to
model the ghost image and to then subtract a scaled model from flat field
images. This method was overall successful, but was prone to adding noise and
leaving sharp residual errors. It remains desirable to avoid the pupil ghost
formation in the first place.

The pupil ghost was found to be greatly suppressed by the ODI
shutter\footnote{ODI uses a Bonn Shutter, http://www.bonn-shutter.de} when the
exposure time was very short (less than 0.1 seconds), and the shutter's blades
would obstruct most of the optical path in front of the dewar window; only a
narrow slit then moves  over the focal plane. This suppression mechanism is
illustrated by Figure \ref{fig_pupilghost}, left. Despite the shutter forming
only a narrow slit, the direct illumination of the focal plane is identical to a
flat field where the shutter completely opens.

This ghost-suppressing behavior of the shutter has been exploited already by
using very short exposure times of order of 50ms for sky flat fields, obsoleting
the need to remove a ghost model from flat fields. But this method is limited to
cases where enough light is available for flat fielding; narrow band filters
cannot be easily calibrated this way.

We have circumvented the limitation to very short calibration exposure times by
drastically lowering the travel speed of the shutter blades during flat field
exposures. The travel time of the shutter blades across the focal plane is
increased to approximately  27 seconds from the factory default of 0.75 seconds.
Now a narrow slot slowly crosses the field of view, allowing effective exposure
times of several seconds.

A good demonstration of the pupil ghost suppression by the shutter is to divide
a flat field where the shutter is operated conventionally by a flat field where
the shutter  operated in the slow narrow-slit mode. Only the extra pupil ghost
component remains visible in the image (Figure \ref{fig_pupilghost}).

All dome flat fields are now acquired using the slow shutter mode, which has
eliminated the need to model and subtract the pupil ghost and other more complex
instrument-internal reflections from flat field calibration images.

\section{Condensation}

A major operational limitation of ODI is the susceptibility of the dewar window
to condensation. As the condensation itself will evaporate quickly under dryer
conditions, we have found that, inevitably, a significant thin film of residue
was left behind, following a condensation event, that could only be removed by
cleaning the dewar window (specifically, drag wiping the surface with alcohol).
The effect of the residue is detrimental for any scientific use of ODI, since it
causes significant scattering and halos around objects, where up to 50\%  of the
light would be lost to scattering.

While the surface of the dewar window is constantly fed dry air to prevent
condensation, we have identified several scenarios in which the condensation  on
the window is insufficiently controlled:

\begin{enumerate}

\item {\bf The ambient humidity is above 70\%}. The ODI instrument was not
designed as an airtight, or slightly over pressurized instrument, and humid,
ambient air can mix into the instrument cavity, the space between dewar window
and front lens, where the filter arms, shutter, and atmospheric dispersion
compensator are located. It was in fact intended that ambient air was actively
drawn through	the instrument cavity in order to improve thermal equalization.

When the shutter remains closed, the injected dry air is confined to the small
space between the dewar window and the shutter. Once the shutter is opened, the 
humid air of the instrument cavity mixes in, and condensation can form on the
cold dewar window. Condensation can appear within a minute after the shutter
(located directly in front of the dewar window) is opened.

As immediate mitigation against condensation, operations of ODI were limited to
ambient humidity below 70\%RH for much of 2017, pending further improvement of
the dry air system (described below); the forced flow of ambient air through the
instrument cavity was inhibited by keeping the shutter closed above the 70\% RH
limit.

\item {\bf Failure of the dry air supply:} The dry air for ODI is generated by a
dehumidifier. At several instances the dry air supply has failed. The failures
are mostly due to a problem in the supplying air compressor (contactor failure
or power outages), but also due to a failure of the dehumidifier, or due to an
interruption in the dry air supply line.

In order to protect against failures in the dry air supply system, a redundant
air compressor and a larger dry air reservoir tank were installed at the
telescope site. The volume and pressure of dry air was increased by a factor of
5 to approximately 5 cfm. Other improvements include the addition of filters
that ensure that the dry air is clean down to a particulate size of 0.1 microns.
Additionally, a gaseous nitrogen tank was added to the dry air system with a
pressure sensitive valve that opens in the event of a reduction in the dry air
flow rate. Due to the large volume of air required to prevent condensation, the
nitrogen handover requires prompt closure of the shutter until the regular air
supply is restored. The humidity and flow of the dry air are actively monitored
with an exception alarm system.

\end{enumerate}

Long term, a redundant method of condensation prevention is desirable for
ODI, e.g., a dewar window heater (as, e.g., implemented for the MMT 
Megacam\cite{McLeod2015}) that would provide of the
order of 60 W power. Heating strips that can be glued to the perimeter of the
dewar window are readily available, and the existing dewar window mount allows
for the addition of heaters with a modest machining effort.

\section{New filters for ODI}

The past two years have seen the addition of new full-frame filters for ODI.  In
addition to the aforementioned g', r', i', and z' broadband filters, we have
procured a u' filter, which has seen regular use for the past few semesters, as
well as a narrow band H$\alpha$ filter which is scheduled to be installed in the
summer of 2018. We have also procured (or are in the process of procuring) four
additional full-frame narrow band filters (422, 659, 695, and 746 nm) which
satisfy the needs of various research projects currently underway using ODI.

\section{Summary}

With the upgrade of the focal plane and the filter drives in 2015, the ODI
instrument has reached a stable state and is a scientifically useful instrument
for the WIYN community\cite{janesh2015, adams2015a, adams2015b, cannon2015, davis2015,
janowiecki2015, janowiecki2015PhDT, janesh2017, jewitt2017, leisman2017, rhode2017, 
wittmann2017, lee2018a, lee2018b, lee2018c,   gorsuch2018 }. ODI is typically 
scheduled for 30 to 35 nights per
semester during dark and grey time. Observers operate the instrument during
their allocated time either locally on site, or via remote observing. The
diverse science observing program includes solar system asteroid follow up
observations, the search for planets via microlensing and occultation, the study of stellar
populations in the  Milky Way star clusters, the evolution of dwarf galaxies
in the Local Group, the search for  counterparts of radio-detected galaxies, and
the hunt for high-redshift galaxies via identifying emission lines in narrow
band filter images.

Condensation residue on the dewar window has been addressed by updating the dry
air supply system, and a redundant  mitigation strategy by installing a heater
has been identified.

The choice to use  OTA detectors in ODI has significantly complicated the
design, construction, and operation of the focal plane, including downstream
data calibration. However, no benefits are derived from the use of OTA
detectors.  A change to conventional CCD detectors, such as STA1600 10kx10k
devices would be the next logical step in the evolution of ODI.

\bibliography{odi} 
\bibliographystyle{spiejour}

\end{document}